\documentclass[aps,prb,twocolumn,showpacs]{revtex4} 
\usepackage{bm}
\usepackage{graphicx}

\begin{document}

\title{Microwave surface resistance in superconductors with grain boundaries}
\author{Yasunori Mawatari}
% \email[]{y.mawatari@aist.go.jp}
\affiliation{%
   Energy Technology Research Institute, 
   National Institute of Advanced Industrial Science and Technology, \\
   Tsukuba, Ibaraki 305--8568, Japan }

\date{Dec. 28, 2004}
% \date{July 27, 2004}

\begin{abstract}
Microwave-field distribution, dissipation, and surface impedance are 
theoretically investigated for superconductors with laminar grain boundaries 
(GBs). 
In the present theory we adopt the two-fluid model for intragrain transport 
current in the grains, and the Josephson-junction model for intergrain 
tunneling current across GBs. 
Results show that the surface resistance $R_s$ nonmonotonically depends 
on the critical current density $J_{cj}$ at GB junctions, and $R_s$ for 
superconductors with GBs can be smaller than the surface resistance 
$R_{s0}$ for ideal homogeneous superconductors without GBs. 
\end{abstract}

% 74.20.De  Phenomenological theories (two-fluid, Ginzburg-Landau, etc.) 
% 74.25.Nf  Response to electromagnetic fields
% 74.50.+r  Tunneling phenomena; point contacts, weak links, Josephson effects
% 74.81.-g  Inhomogeneous superconductors and superconducting systems 

\maketitle

\section{Introduction} %*****
High-temperature superconductors contain many grain boundaries (GBs), 
where the order parameter is locally suppressed due to the short coherence 
length.~\cite{Deutscher87} 
GBs have attracted much interest for their basic physics as well as 
for their applications in 
superconductors,~\cite{Mannhart99,Larbalestier01,Hilgenkamp02} 
and play a crucial role in microwave response and surface resistance 
$R_s$ of high-temperature superconducting 
films.~\cite{Hylton88,Attanasio91,Halbritter92,Nguyen93,Fagerberg94,Mahel96,McDonald97,Gallop97,Obara01} 

Electrodynamics of GB junctions can be described using the Josephson-junction 
model, and one of the most important parameters that characterize GB junctions 
is the critical current density $J_{cj}$ for Josephson tunneling current 
across GBs.~\cite{Barone82,Tinkham96,VanDuzer99} 
The $J_{cj}$ strongly depends on the misorientation angle of 
GBs.~\cite{Dimos88,Gurevich98} 
In $\rm YBa_2Cu_3O_{7-\delta}$ films, $J_{cj}$ can be 
enhanced~\cite{Hammerl00} and $R_s$ reduced~\cite{Obara01} by Ca doping. 
The investigation of the relationship between $R_s$ and $J_{cj}$ is 
needed to understand the behavior of $R_s$ and $J_{cj}$ in Ca doped 
$\rm YBa_2Cu_3O_{7-\delta}$ films. 
The $J_{cj}$ dependence of $R_s$, however, has not yet been clarified, 
and it is not trivial whether GBs enhance the microwave dissipation 
that is proportional to $R_s$. 

In this paper, we present theoretical investigation on the microwave 
field and dissipation in superconductors with laminar GBs. 
Theoretical expressions of the surface impedance $Z_s=R_s-iX_s$ of 
superconductors with GBs are derived as functions of $J_{cj}$ at GB junctions.

\section{Basic Equations} %*****
\subsection{Superconductors with grain boundaries} %*****
We consider penetration of a microwave field (i.e., magnetic induction 
$\bm{B}=\mu_0\bm{H}$, electric field $\bm{E}$, and current density 
$\bm{J}$) into superconductors that occupy a semi-infinite area of $x>0$. 
We investigate linear response for small microwave power limit, such that 
the time dependence of the microwave field is expressed by the harmonic 
factor, $e^{-i\omega t}$, where $\omega/2\pi$ is the microwave frequency 
that is much smaller than the energy-gap frequency of the superconductors. 
Magnetic induction $\bm{B}$ is assumed to be less than the lower critical 
field, such that no vortices are present in the superconductors. 
(See Ref.~\onlinecite{Coffey91} for microwave response of vortices.) 

The GBs are modelled to have laminar structures as in 
Ref.~\onlinecite{Hylton89}; the laminar GBs that are parallel to the $xz$ 
plane are situated at $y=ma$, where $a$ is the spacing between grains 
(i.e., effective grain size) and $m=0,\pm 1,\pm 2,\cdots, \pm\infty$. 
The thickness of the barrier of GB junctions, $d_j$, is much smaller than 
both $a$ and the London penetration depth $\lambda$, and therefore, 
we investigate the thin-barrier limit of $d_j\to 0$, namely, GB barriers 
situated at $ma-0<y<ma+0$.

\subsection{Two-fluid model for intragrain current} %*****
We adopt the standard two-fluid model~\cite{Tinkham96,VanDuzer99} for 
current transport in the grain at $ma+0<y<(m+1)a-0$. 
The intragrain current $\bm{J}=\bm{J}_s+\bm{J}_n$ is given by the sum of 
the supercurrent $\bm{J}_s=i\sigma_s\bm{E}$ and the normal current 
$\bm{J}_n=\sigma_n\bm{E}$, where $\sigma_s=1/\omega\mu_0\lambda^2$ and 
$\sigma_n$ is the normal-fluid conductivity in the grains. 
The displacement current $\bm{J}_d=-i\omega\epsilon\bm{E}$ with the 
dielectric constant $\epsilon$ can be neglected for a microwave range of 
$\omega/2\pi\sim$ GHz. 
Amp\`{e}re's law $\mu_0^{-1}\nabla\times\bm{B}=(\sigma_n+i\sigma_s)\bm{E}$ 
is thus reduced to 
\begin{equation}
   \bm{E}= -i\omega\Lambda_g^2 \nabla\times\bm{B} , 
\label{E_intra}
\end{equation}
where $\Lambda_g$ is the intragrain ac field penetration depth defined by 
\begin{equation}
   \Lambda_g^{-2} =\omega\mu_0(\sigma_s -i\sigma_n) 
   =\lambda^{-2} -i\omega\mu_0\sigma_n . 
\label{Lg_intra}
\end{equation}
Combining Eq.~(\ref{E_intra}) with Faraday's law, 
$\nabla\times\bm{E} =i\omega\bm{B}$, we obtain the London equation 
for magnetic induction $\bm{B}=B_z(x,y)\hat{\bm{z}}$ for $y\neq ma$ as 
\begin{equation}
   B_z -\Lambda_g^2\nabla^2 B_z =0 . 
\label{B_intra}
\end{equation}

For ideal homogeneous superconductors without GBs, Eq.~(\ref{B_intra}) 
is valid for $-\infty<y<+\infty$ and the solution is simply given by 
$B_z(x)= \mu_0 H_0e^{-x/\Lambda_g}$, and the electric field is obtained 
from Eq.~(\ref{E_intra}) as $E_y(x)=-i\omega\mu_0\Lambda_g H_0e^{-x/\Lambda_g}$. 
The surface impedance $Z_{s0}=R_{s0}-iX_{s0}$ for homogeneous superconductors 
is given by $Z_{s0}=E_y(x=0)/H_0= -i\omega\mu_0\Lambda_g$. 
The surface resistance $R_{s0}=\mbox{Re}(Z_{s0})$ and reactance 
$X_{s0}=-\mbox{Im}(Z_{s0})$ of ideal homogeneous superconductors without 
GBs are given by~\cite{VanDuzer99} 
\begin{eqnarray}
   R_{s0} &=& \mu_0^2\omega^2\lambda^3\sigma_n/2 , 
\label{Rs0_two-fluid}\\
   X_{s0} &=& \mu_0\omega\lambda 
\label{Xs0_two-fluid}
\end{eqnarray}
for $\sigma_n/\sigma_s\ll 1$ well below the superconducting transition 
temperature $T_c$. 

\subsection{Josephson-junction model for intergrain current} %*****
We adopt the Josephson-junction model~\cite{Barone82,Tinkham96,VanDuzer99} 
for tunneling current across GBs at $y=ma$. 
Behavior of the GB junctions is determined by the gauge-invariant phase 
difference across GBs, $\varphi_j(x)$, and the voltage induced across GB, 
$V_j(x)$, is given by the Josephson's relation, 
\begin{equation}
   \int_{ma-0}^{ma+0} E_y dy 
   = V_j = \frac{\phi_0}{2\pi}(-i\omega\varphi_j) , 
\label{Vj_GB}
\end{equation}
where $\phi_0$ is the flux quantum. 
The tunneling current parallel to the $y$ axis is given by the sum of the 
superconducting tunneling current (i.e., Josephson current) 
$J_{sj}=J_{cj}\sin\varphi_j$ and the normal tunneling current 
(i.e., quasiparticle tunneling current) $J_{nj}=\gamma_{nj}V_j$. 
The critical current density $J_{cj}$ at GB junctions is one of the most 
important parameters in the present paper, and the resistance-area product 
of GB junctions corresponds to $1/\gamma_{nj}$. 
We neglect the displacement current across GBs, $J_{dj}=-i\omega C_jV_j$ 
where $C_j$ is the capacitance of the GB junctions. 

Here we define the Josephson length $\lambda_J$ and the characteristic 
current density $J_0$ as \begin{eqnarray}
   \lambda_J &=& (\phi_0/4\pi\mu_0 J_{cj}\lambda)^{1/2} , 
\label{Lambda-J}\\
   J_0 &=& \phi_0/4\pi\mu_0\lambda^3 . 
\label{J0}
\end{eqnarray}
The ratio $J_{cj}/J_0=(\lambda/\lambda_J)^2$ characterizes the coupling 
strength of GB junctions.~\cite{Gurevich92} 
For weakly coupled GBs, namely, $J_{cj}/J_0=(\lambda/\lambda_J)^2\ll 1$ 
(e.g., high-angle GBs), electrodynamics of the GB junctions can be well 
described by the weak-link model.~\cite{Barone82,Tinkham96,VanDuzer99} 
For strongly coupled GBs, namely, $J_{cj}/J_0=(\lambda/\lambda_J)^2\agt 1$ 
(e.g., low-angle GBs), the Josephson-junction model is still valid but 
requires appropriate boundary condition at GBs, as given in Eq.~(4) in 
Ref.~\onlinecite{Gurevich92}, as pointed out by Gurevich; 
see also Refs.~\onlinecite{Hylton89} and \onlinecite{Ivanchenko90}. 

In the small-microwave-power limit such that 
$\sin\varphi_j\simeq \varphi_j=2\pi V_j/(-i\omega\phi_0)$ for 
$|\varphi_j|\ll 1$, the $J_{cj}$ is reduced to 
\begin{equation}
   J_{sj}\simeq J_{cj}\varphi_j =i\gamma_{sj}V_j , 
\label{J-gb_super-linear}
\end{equation}
where $\gamma_{sj}= 2\pi J_{cj}/\omega\phi_0 
=1/2\omega\mu_0\lambda\lambda_J^2$ . 
The total tunneling current across GB is thus given by 
\begin{equation}
   \left. -\frac{1}{\mu_0} \frac{\partial B_z}{\partial x} \right|_{y=ma} 
   = J_{sj}+J_{nj} = (i\gamma_{sj} +\gamma_{nj}) V_j . 
\label{J-V_inter-sigma}
\end{equation}
Integration of Faraday's law, 
$\partial E_y/\partial x -\partial E_x/\partial y=i\omega B_z$, yields 
\begin{eqnarray}
   \lefteqn{E_x(x,y=ma+0) -E_x(x,y=ma-0)} &&
\nonumber\\
   &=& \int_{ma-0}^{ma+0}dy \left[ 
      \frac{\partial E_y(x,y)}{\partial x} -i\omega B_z(x,y) \right] 
   = \frac{\partial V_j(x)}{\partial x} , \quad
\label{Ex_discontinue}
\end{eqnarray}
where we used Eq.~(\ref{Vj_GB}). 
The static version (i.e., $\omega\to 0$) of Eq.~(\ref{Ex_discontinue}) 
corresponds to Eq.~(4) in Ref.~\onlinecite{Gurevich92}. 
Substitution of Eqs.~(\ref{E_intra}) and (\ref{J-V_inter-sigma}) into 
Eq.~(\ref{Ex_discontinue}) yields the boundary condition for $B_z$ at $y=ma$, 
\begin{equation}
   \left. -\frac{\partial B_z}{\partial y} \right|_{y=ma+0} 
      \left. +\frac{\partial B_z}{\partial y} \right|_{y=ma-0} 
   = \left. \frac{a\Lambda_j^2}{\Lambda_g^2} 
      \frac{\partial^2 B_z}{\partial x^2} \right|_{y=ma} , 
\label{bc-GB}
\end{equation}
where $\Lambda_j$ is the characteristic length for ac field penetration 
into GBs defined by 
\begin{eqnarray}
   \Lambda_j^{-2} &=& \omega\mu_0 a(\gamma_{sj} -i\gamma_{nj}) 
\nonumber\\
   &=& \mu_0 a\left( 2\pi J_{cj}/\phi_0 
      -i\omega\gamma_{nj}\right) . 
\label{Lj_gb}
\end{eqnarray}

\section{Surface impedance} %*****
\subsection{Microwave field and surface impedance}
Equations~(\ref{B_intra}) and (\ref{bc-GB}) are combined into a single 
equation for $x>0$ and $-\infty<y<+\infty$ as 
\begin{equation}
   B_z -\Lambda_g^2 \nabla^2B_z 
   = a\Lambda_j^2 \sum_{m=-\infty}^{+\infty} 
      \frac{\partial^2 B_z}{\partial x^2} \delta(y-ma) , 
\label{B_all}
\end{equation}
whose solution is calculated as 
\begin{eqnarray}
   \frac{B_z(x,y)}{\mu_0H_0} 
   &=& e^{-x/\Lambda_g} 
      +\frac{2}{\pi}\int_0^{\infty}dk\, 
      \frac{\cosh[K(y-a/2)]}{\Lambda_g^{\,2}K^2\sinh(Ka/2)} 
\nonumber\\
   && \times \frac{k\sin kx}{(2K\Lambda_g^2/a\Lambda_j^2) +k^2\coth(Ka/2)} 
\label{Bz_0<y<a}
\end{eqnarray}
for $0<y<a$, where $K=(k^2+\Lambda_g^{-2})^{1/2}$. 
The right-hand side of Eq.~(\ref{B_all}) and the second term 
of the right-hand side of Eq.~(\ref{Bz_0<y<a}) reflect the GB effects. 
See Appendix A for the derivation of Eq.~(\ref{Bz_0<y<a}) from 
Eq.~(\ref{B_all}). 

Electric field in the grains is obtained from Eq.~(\ref{E_intra}) as 
$E_y= i\omega\Lambda_g^{\,2}\partial B_z/\partial x$, and voltage induced 
across GB is obtained from Eq.~(\ref{J-V_inter-sigma}) as 
$V_j=\left. i\omega a\Lambda_j^2\partial B_z/\partial x\right|_{y=0}$. 
The mean electric field $\bar{E}_s$ at the surface of the superconductor 
is thus calculated as 
\begin{eqnarray}
   \bar{E}_s &\equiv& 
      \frac{1}{a} \int_{-0}^{a-0}dy\, E_y(x=0,y) 
\nonumber\\
   &=& \frac{1}{a} \left[ V_j(x=0) +\int_{+0}^{a-0}dy\, E_y(x=0,y) \right] 
\nonumber\\
   &=& i\omega \left[ \Lambda_j^2 
      \left. \frac{\partial B_z}{\partial x}\right|_{x=y=0} 
      + \frac{\Lambda_g^{\,2}}{a}\int_{+0}^{a-0} dy\, 
      \left. \frac{\partial B_z}{\partial x}\right|_{x=0} \right] . 
\nonumber\\
\label{mean-Es_def}
\end{eqnarray}
Substitution of Eq.~(\ref{Bz_0<y<a}) into Eq.~(\ref{mean-Es_def}) yields 
the surface impedance $Z_s=R_s-iX_s \equiv\bar{E}_s/H_0$ as 
\begin{eqnarray}
   \frac{Z_s}{-i\omega\mu_0\Lambda_g}
   &=& 1+\frac{2}{\pi} \int_0^{\infty} dk \frac{1}{\Lambda_g^{\,3}K^3} 
\nonumber\\
   && \times \frac{1}{(K\Lambda_g^2/\Lambda_j^2) +(k^2a/2)\coth(Ka/2)} . 
   \qquad
\label{Zs_general}
\end{eqnarray}
The surface resistance and reactance are given by $R_s= \mbox{Re}(Z_s)$ 
and $X_s= -\mbox{Im}(Z_s)$, respectively. 

\subsection{Microwave dissipation and surface resistance}
The time-averaged electromagnetic energy passing through the surface of 
a superconductor at $x=0$ and $-0<y<a-0$ is given by the real part of 
\begin{equation}
   {\cal E}= \frac{1}{2\mu_0}\int_{-0}^{a-0}dy (E_yB_z^*)_{x=0} 
   = \frac{a}{2}\bar{E}_sH_0^* , 
\label{Poyntings-vector}
\end{equation}
where $\bar{E}_s=Z_sH_0$ is defined by Eq.~(\ref{mean-Es_def}), 
and $(B_z)_{x=0}=\mu_0 H_0$. 
Poynting's theorem~\cite{Jackson75} states that $\cal E$ is identical 
to the energy stored and dissipated in the superconductor, 
\begin{eqnarray}
   {\cal E}
   &=& \frac{1}{2} 
      \int_0^{\infty}dx\left[ 
      \int_{+0}^{a-0}dy\, (\sigma_n -i\sigma_s)|\bm{E}|^2 \right.
\nonumber\\
   && \left.\mbox{} +(\gamma_{nj}-i\gamma_{sj})|V_j|^2 
      -\int_{-0}^{a-0}dy\, \frac{i\omega}{\mu_0}|B_z|^2 \right] . 
\label{Poyntings-theorem}
\end{eqnarray}
The real parts of Eqs.~(\ref{Poyntings-vector}) and (\ref{Poyntings-theorem}) 
show that the surface resistance $R_s=\mbox{Re}(\bar{E}_s/H_0)=\mbox{Re}(Z_s)$ 
is composed of two terms: 
\begin{equation}
   R_s = R_{sg} +R_{sj} . 
\label{Rs_Rsg+Rsj}\\
\end{equation}
The intragrain contribution $R_{sg}$ is from the energy dissipation 
in the grains, and the intergrain contribution $R_{sj}$ is from the 
dissipation at GBs: 
\begin{eqnarray}
   R_{sg} &=& 
      \frac{1}{a|H_0|^2} 
      \int_0^{\infty}dx\int_{+0}^{a-0}dy\, \sigma_n|\bm{E}|^2 , 
\label{Rsg_dissipation}\\
   R_{sj} &=& 
      \frac{1}{a|H_0|^2} \int_0^{\infty}dx\, \gamma_{nj}|V_j|^2 . 
\label{Rsj_dissipation}
\end{eqnarray}

Both the intragrain current $|\bm{J}_g|$ around GBs and the intergrain 
tunneling current $|J_j|$ across GBs are suppressed by the GBs, and are 
increasing functions of $J_{cj}$. 
With increasing $J_{cj}$, the intragrain electric field 
$|\bm{E}|=|\bm{J}_g/(\sigma_n+i\sigma_s)|$ also increases, whereas the 
intergrain voltage $|V_j|=|J_j/(\gamma_{nj}+i\gamma_{sj})|$ decreases 
because $\gamma_{sj}\propto J_{cj}$. 
The dissipation in the grains, $\sigma_n|\bm{E}|^2/2$, and the intragrain 
contribution to the surface resistance, $R_{sg}$, therefore, 
tend to {\em increase} with increasing $J_{cj}$. 
The dissipation at GBs, $\gamma_{nj}|V_j|^2/2$, and the intergrain 
contribution to the surface resistance, $R_{sj}$, on the other hand, 
{\em decrease} with increasing $J_{cj}$. 

The surface reactance $X_s=-\mbox{Im}(Z_s)$ is also divided into two 
contributions, 
\begin{equation}
   X_s = X_{sg} +X_{sj} , 
\label{Xs_Xsg+Xsj}\\
\end{equation}
where the intragrain contribution $X_{sg}$ and the intergrain 
contribution $X_{sj}$ 
are given by 
\begin{eqnarray}
   X_{sg} &=& \frac{1}{a|H_0|^2} 
      \int_0^{\infty}dx\int_{+0}^{a-0}dy\, \left( \sigma_s|\bm{E}|^2 
      +\frac{\omega}{\mu_0}|B_z|^2 \right) , 
\nonumber\\
\label{Xsg_energy}\\
   X_{sj} &=& \frac{1}{a|H_0|^2} \int_0^{\infty}dx\, \gamma_{sj}|V_j|^2 . 
\label{Xsj_energy}
\end{eqnarray}
Both $X_{sg}$ and $X_{sj}$ decrease with increasing $J_{cj}$.

\subsection{Simplified expressions for surface impedance} %*****
The following Eqs.~(\ref{Zs_small-grain})--(\ref{Xs_large-Jc}) show 
simplified expressions of the surface impedance $Z_s$, the surface 
resistance $R_s=\mbox{Re}(Z_s)$, and the surface reactance 
$X_s=-\mbox{Im}(Z_s)$ for certain restricted cases, assuming 
$\sigma_n/\sigma_s\ll 1$ and $\gamma_{nj}/\gamma_{sj}\ll 1$ well below 
the transition temperature. 

For small grains of $a\ll\lambda$ such that $\coth(Ka/2)\simeq 2/Ka$, 
Eq.~(\ref{Zs_general}) is reduced to 
\begin{equation}
   Z_s\simeq -i\omega\mu_0 \left(\Lambda_g^2 +\Lambda_j^2\right)^{1/2} . 
\label{Zs_small-grain}
\end{equation}
The right-hand side of Eq.~(\ref{B_all}) is reduced to 
$\Lambda_j^2\partial^2B_z/\partial x^2$ for $a\ll\lambda$, and the 
effective ac penetration depth is given by 
$\Lambda_{\rm eff}= (\Lambda_g^2 +\Lambda_j^2)^{1/2}$ as in 
Ref.~\onlinecite{Hylton89}, resulting in the surface impedance given 
by Eq.~(\ref{Zs_small-grain}). 
The $R_s$ and $X_s$ for small grains is obtained as 
\begin{eqnarray}
   \frac{R_s}{R_{s0}} &\simeq& 
      \left(1+\frac{2\lambda}{a}\frac{J_0}{J_{cj}}\right)^{-1/2} 
      \left[\,1 +\frac{4\lambda^2\gamma_{nj}}{a\sigma_n} 
      \left(\frac{J_0}{J_{cj}}\right)^2 \,\right] , \qquad
\label{Rs_small-grain}\\
   \frac{X_s}{X_{s0}} &\simeq& 
      \left(1+\frac{2\lambda}{a}\frac{J_0}{J_{cj}}\right)^{+1/2} , 
\label{Xs_small-grain}
\end{eqnarray}
where $R_{s0}$, $X_{s0}$, and $J_0$ are defined by 
Eqs.~(\ref{Rs0_two-fluid}), (\ref{Xs0_two-fluid}) and (\ref{J0}), respectively. 
Equation~(\ref{Rs_small-grain}) is decomposed into the intragrain 
$R_{sg}$ and intergrain $R_{sj}$ contributions, 
as $R_{sg}/R_{s0}\simeq (1+2\lambda J_0/aJ_{cj})^{-1/2}$ and 
$R_{sj}/R_{sg}\simeq (4\lambda^2\gamma_{nj}/a\sigma_n) (J_0/J_{cj})^2$, 
respectively. 

Equation~(\ref{Zs_small-grain}) is further simplified when 
$a\ll 2\lambda_J^{\,2}/\lambda$ for small grain and weakly coupled GBs as 
\begin{equation}
   Z_s \simeq -i\omega\mu_0\Lambda_j , 
\label{Zs_small-grain-weak-link}
\end{equation}
and we have 
\begin{eqnarray}
   \frac{R_s}{R_{s0}} &\simeq& 
      \frac{2\gamma_{nj}\lambda}{\sigma_n} 
      \left(\frac{2\lambda}{a}\right)^{1/2} 
      \left(\frac{J_0}{J_{cj}}\right)^{3/2} , 
\label{Rs_small-grain-weak-link}\\
   \frac{X_s}{X_{s0}} &\simeq& 
      \left(\frac{2\lambda}{a}\right)^{1/2} 
      \left(\frac{J_0}{J_{cj}}\right)^{1/2} . 
\label{Xs_small-grain-weak-link}
\end{eqnarray}
Thus, we obtain the dependence of $R_s$ and $X_s$ on the material 
parameters as $R_s\propto \gamma_{nj}a^{-1/2}J_{cj}^{-3/2}$ and 
$X_s\propto a^{-1/2}J_{cj}^{-1/2}$, which are independent of $\lambda$. 
The $R_s$ given by Eq.~(\ref{Rs_small-grain-weak-link}) for the small 
grain and weakly coupled GBs is mostly caused by intergrain dissipation, 
$R_s\simeq R_{sj}\gg R_{sg}$. 
For $X_s$ given by Eq.~(\ref{Xs_small-grain-weak-link}), on the other hand, 
both intragrain $X_{sg}$ and intergrain $X_{sj}$ contribute to the total 
$X_s=X_{sg}+X_{sj}$.

For large $J_{cj}$ (i.e., strong-coupling limit) such that 
$K\Lambda_g^2/\Lambda_j^2\gg (k^2a/2)\coth(Ka/2)$, Eq.~(\ref{Zs_general}) 
for the surface impedance $Z_s$ is simplified as 
\begin{equation}
   Z_s \simeq -i\omega\mu_0 \left(\Lambda_g +\Lambda_j^2/2\Lambda_g\right) , 
\label{Zs_large-Jcj}
\end{equation}
and we have 
\begin{eqnarray}
   \frac{R_s}{R_{s0}} &\simeq& 
      1 -\frac{\lambda}{a}\frac{J_0}{J_{cj}} 
      +\frac{4\lambda^2\gamma_{nj}}{a\sigma_n} 
      \left(\frac{J_0}{J_{cj}}\right)^2 , 
\label{Rs_large-Jc}\\
   \frac{X_s}{X_{s0}} &\simeq& 
      1 +\frac{\lambda}{a}\frac{J_0}{J_{cj}} . 
\label{Xs_large-Jc}
\end{eqnarray}
The first and second terms of the right-hand side of Eq.~(\ref{Rs_large-Jc}) 
correspond to the intragrain contribution, $R_{sg}$, whereas the third term 
corresponds to the intergrain contribution, $R_{sj}$.

\section{Discussion} %*****
% \section{Dependence of $R_s$ on $J_{cj}$} %*****

\begin{figure}[b,t,h]%*************
\includegraphics{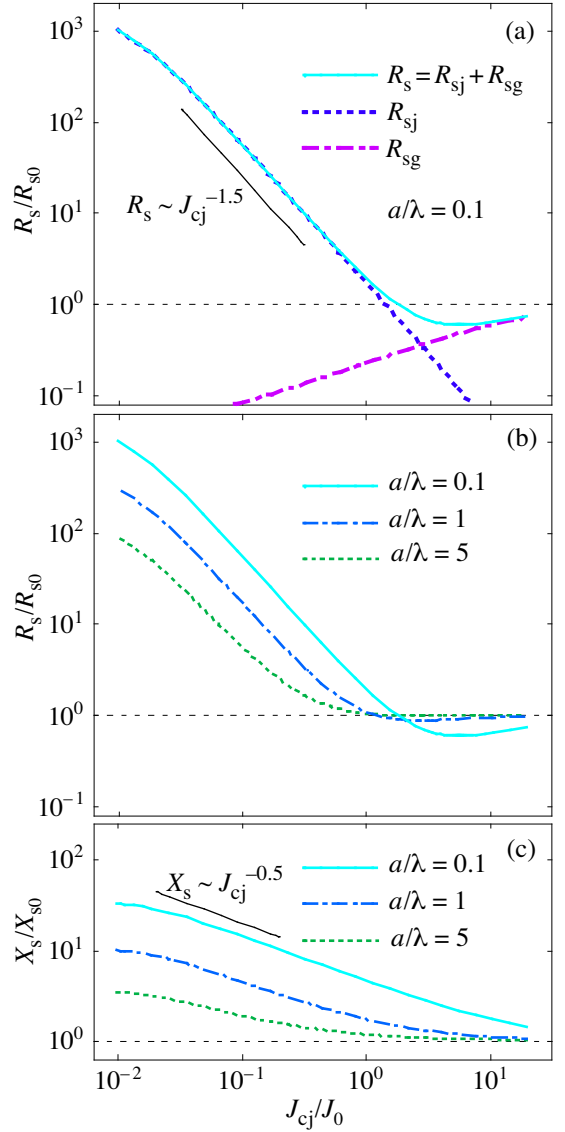}
\caption{%
Dependence of surface resistance $R_s=\mbox{Re}(Z_s)$ and surface 
reactance $X_s=-\mbox{Im}(Z_s)$ [i.e., Eq.~(\ref{Zs_general}) with 
Eqs.~(\ref{Lg_intra}) and (\ref{Lj_gb})] on critical current density 
$J_{cj}$ at GB junctions. 
$R_s$ is normalized to the surface resistance without GB, i.e., $R_{s0}$ 
given by Eq.~(\ref{Rs0_two-fluid}), $X_s$ is normalized to $X_{s0}$ given 
by Eq.~(\ref{Xs0_two-fluid}), and $J_{cj}$ is normalized to $J_0$ defined 
as Eq.~(\ref{J0}). 
Parameters are $\omega/2\pi=10\,$GHz, $\lambda=0.2\,\mu$m, 
$\sigma_n=10^7\,\Omega^{-1}$m$^{-1}$, and 
$\gamma_{nj}=10^{13}\,\Omega^{-1}$m$^{-2}$, which yield 
$R_{s0}=0.25\,$m$\Omega$, $X_{s0}=16\,$m$\Omega$, and 
$J_0=1.6\times 10^{10}\,$A/m$^2$. 
(a) Total surface resistance $R_s=R_{sj}+R_{sg}$, 
intergrain contribution $R_{sj}$ given by Eq.~(\ref{Rsj_dissipation}), 
and intragrain contribution $R_{sg}$ given by 
Eq.~(\ref{Rsg_dissipation}) for $a/\lambda=0.1$. 
(b) $R_s$ and (c) $X_s$ for $a/\lambda=0.1$, $1$, and $5$.
}
\label{Fig_Rs-Jc}
\end{figure}

Figure~\ref{Fig_Rs-Jc}(a) and (b) shows $J_{cj}$ dependence of $R_s$. 
As shown in Fig.~\ref{Fig_Rs-Jc}(a), the intergrain contribution $R_{sj}$ 
is dominant for weakly coupled GBs (i.e., small $J_{cj}/J_0$ regime), 
whereas the intragrain contribution $R_{sg}$ is dominant for strongly 
coupled GBs (i.e., large $J_{cj}/J_0$). 
The $R_{sj}$ {\em decreases} with increasing $J_{cj}$ as 
$R_{sj}\propto J_{cj}^{-1.5}$ [see Eq.~(\ref{Rs_small-grain-weak-link})], 
whereas $R_{sg}$ {\em increases} with $J_{cj}$. 
The resulting surface resistance $R_s=R_{sj}+R_{sg}$ {\em nonmonotonically} 
depends on $J_{cj}$ and has a minimum, because $R_s$ is determined by the 
competition between $R_{sj}$ and $R_{sg}$. 
As shown in Fig.~\ref{Fig_Rs-Jc}(c), on the other hand, $X_s$ monotonically 
decreases with increasing $J_{cj}$ [i.e., $X_s\propto J_{cj}^{-0.5}$ for 
weakly coupled GBs as in Eq.~(\ref{Xs_small-grain-weak-link})]. 

The nonmonotonic dependence of $R_s$ on the grain size $a$ is also seen 
in Fig.~\ref{Fig_Rs-Jc}(b). 
For small $J_{cj}/J_0$ the $R_s$ decreases with increasing $a$ as 
$R_s\propto a^{-0.5}$ [see Eq.~(\ref{Rs_small-grain-weak-link})], 
whereas $R_s$ increases with $a$ for large $J_{cj}/J_0$. 

The $R_s$ for strongly coupled GBs can be {\em smaller} than $R_{s0}$ 
for ideal homogeneous superconductors without GBs, namely, $R_s/R_{s0}<1$ 
for $J_{cj}/J_0\agt 1$. 
The minimum surface resistance for $\lambda\gamma_{nj}/\sigma_n=0.2$ is 
$R_s/R_{s0}\approx 0.97$ for $a/\lambda=5$, $R_s/R_{s0}\approx 0.86$ for 
$a/\lambda=1$, and $R_s/R_{s0}\approx 0.59$ for $a/\lambda=0.1$. 
The minimum $R_s/R_{s0}$ is further reduced when 
$\lambda\gamma_{nj}/\sigma_n$ is further reduced. 

Theoretical results shown above may possibly be observed by measuring $R_s$, 
$X_s$, and $J_{cj}$ in Ca doped $\rm YBa_2Cu_3O_{7-\delta}$ films. 
The enhancement of $J_{cj}$ (Ref.~\onlinecite{Hammerl00}) and reduction of 
$R_s$ (Ref.~\onlinecite{Obara01}) by Ca doping are individually observed 
in $\rm YBa_2Cu_3O_{7-\delta}$, but simultaneous measurements of $J_{cj}$ 
and $R_s$ are needed to investigate the relationship between $R_s$ and 
$J_{cj}$. 
The nonmonotonic $J_{cj}$ dependence of $R_s$ for strongly coupled GBs 
may be observed in high quality films with small grains $a<\lambda$ 
and with large $J_{cj}$ on the order of $J_0\sim 10^{10}\,{\rm A/m}^2$ 
at low temperatures.

\section{Conclusion} %*****
We have theoretically investigated the microwave-field distribution in 
superconductors with laminar GBs. 
The field calculation is based on the two-fluid model for current transport 
in the grains and on the Josephson-junction model for tunneling current 
across GBs. 
Results show that the microwave dissipation at GBs is dominant for weakly 
coupled GBs of $J_{cj}\ll J_0$, whereas dissipation in the grains is 
dominant for strongly coupled GBs of $J_{cj}\gg J_0$. 
The surface resistance $R_s$ nonmonotonically depends on $J_{cj}$; 
the $R_s$ decreases with increasing $J_{cj}$ as $R_s\propto J_{cj}^{-1.5}$ 
for $J_{cj}\ll J_0$, whereas $R_s$ increases with $J_{cj}$ for 
$J_{cj}\gg J_0$. 
The intragrain dissipation can be suppressed by GBs, and the surface 
resistance of superconductors with GBs can be smaller than that of ideal 
homogeneous superconductors without GBs.

% \section{Acknowlegdments} %*****
\begin{acknowledgments}
I gratefully acknowledge H. Obara, J.C. Nie, A. Sawa, M. Murugesan, 
H. Yamasaki, and S. Kosaka for stimulating discussions. 
\end{acknowledgments}

\appendix
\section{} %*****
Equation~(\ref{Bz_0<y<a}) is derived by solving Eq.~(\ref{B_all}) 
with the boundary condition of $B_z=\mu_0H_0$ at $x=0$, as follows. 

We introduce the Fourier transform of $B_z(x,y)$ and $B_z(x,ma)=B_z(x,0)$ as 
\begin{eqnarray}
   \tilde{b}(k,q) &=& \int_0^{\infty}dx \int_{-\infty}^{+\infty}dy\, 
      B_z(x,y) e^{-iqy}\sin kx , 
\label{bkq_Fourier}\\
   \tilde{b}_0(k) &=& \int_0^{\infty}dx\, B_z(x,0) \sin kx 
   = \int_{-\infty}^{+\infty}\frac{dq}{2\pi}\, \tilde{b}(k,q) , 
\nonumber\\
\label{b0k_Fourier}
\end{eqnarray}
respectively. 
The Fourier transform of Eq.~(\ref{B_all}) leads to 
\begin{eqnarray}
   \frac{\tilde{b}(k,q)}{\mu_0H_0} &=& 2\pi\delta(q)\frac{k}{K^2} 
\nonumber\\
   && {}+\frac{\alpha k}{K^2+q^2}\sum_m e^{-imqa} 
      \left[1-\frac{k\tilde{b}_0(k)}{\mu_0H_0}\right] , 
\label{bkq-b0k}
\end{eqnarray}
where $K=(k^2+\Lambda_g^{-2})^{1/2}$ and $\alpha= a\Lambda_j^2/\Lambda_g^2$. 
Substituting Eq.~(\ref{bkq-b0k}) into Eq.~(\ref{b0k_Fourier}), we have 
\begin{eqnarray}
   \frac{\tilde{b}_0(k)}{\mu_0H_0} &=& \frac{k}{K^2} 
      +\alpha k \left[1-\frac{k\tilde{b}_0(k)}{\mu_0H_0}\right] 
      \sum_m \int_{-\infty}^{+\infty}\frac{dq}{2\pi} 
      \frac{e^{-imqa}}{K^2+q^2} , 
\nonumber\\
%  &=& \frac{k}{K^2} 
%     +\alpha k \left[1-\frac{k\tilde{b}_0(k)}{\mu_0H_0}\right] 
%     \frac{1}{2K}\coth\left(\frac{Ka}{2}\right) , 
\label{b0k-cal}
\end{eqnarray}
which is reduced to 
\begin{equation}
   \frac{\tilde{b}_0(k)}{\mu_0H_0} = \frac{1}{k} 
      -\frac{2}{kK \Lambda_g^2} \frac{1}{2K+\alpha k^2\coth(Ka/2)} . 
\label{b0k-sol}
\end{equation}
$B_z(x,y)$ is calculated from $\tilde{b}(k,q)$ given by Eq.~(\ref{bkq-b0k}) as 
\begin{eqnarray}
   \frac{B_z(x,y)}{\mu_0H_0} &=& \frac{2}{\pi}\int_0^{\infty}dk 
      \int_{-\infty}^{+\infty}\frac{dq}{2\pi}\, 
      \frac{\tilde{b}(k,q)}{\mu_0H_0} e^{iqy}\sin kx 
\nonumber\\
   &=& e^{-x/\Lambda_g} 
      +\frac{2\alpha}{\pi} \int_0^{\infty}dk\, 
      k\sin kx \left[1-\frac{k\tilde{b}_0(k)}{\mu_0H_0}\right] 
\nonumber\\
   && {}\times \sum_m \int_{-\infty}^{+\infty}\frac{dq}{2\pi} 
      \frac{e^{iq(y-ma)}}{K^2+q^2} . 
\label{Bz-cal}
\end{eqnarray}
Substitution of Eq.~(\ref{b0k-sol}) into Eq.~(\ref{Bz-cal}) yields 
Eq.~(\ref{Bz_0<y<a}).

\end{document}